\documentclass[twocolumn,preprintnumbers,amsmath,amssymb]{revtex4}

\usepackage{graphicx}
\usepackage{dcolumn}
\usepackage{bm,epsfig}

\begin{document}

\title{Unveiling the hybridization gap in Ce$_2$RhIn$_8$ heavy fermion compound}

\author{C. Adriano$^{1,2}$} 
\author{F. Rodolakis$^{3}$} \altaffiliation{Currently at Northwestern University Argonne National Laboratory Institute of Science and Engineering (NAISE), Northwestern University, Evanston, Illinois 60208, USA} 
\author{P. F. S. Rosa$^{1}$} 
\author{F. Restrepo$^{2}$}
\author{M. A. Continentino$^{4}$}
\author{Z. Fisk$^{5}$} 
\author{J. C. Campuzano$^{2,3}$}
\author{P. G. Pagliuso$^{1}$}

\affiliation{$^{1}$Instituto de F\'isica \lq\lq Gleb Wataghin\rq\rq,
UNICAMP, Campinas-SP, 13083-970, Brazil.\\
$^{2}$Department of Physics, University of Illinois at Chicago,
Chicago,
Illinois 60607, USA.\\
$^{3}$Formerly at the Materials Science Division, Argonne National Laboratory,
Argonne, Illinois 60439, USA.\\
$^{4}$Centro Brasileiro de Pesquisas F\'isicas, Rua Dr. Xavier
Sigaud
150, 22290-180, Rio de Janeiro, RJ, Brazil.\\
$^{5}$ University of California, Irvine, California 92697-4574,USA}

\date{\today}

\begin{abstract}
A Kondo lattice of strongly interacting $f$-electrons immersed in a sea of conduction electrons remains one of the unsolved problems in condensed matter physics. The problem concerns localized $f$-electrons at high temperatures which evolve into hybridized heavy quasi-particles at low temperatures, resulting in the appearance of a hybridization gap. Here, we unveil the presence of hybridization gap in Ce$_2$RhIn$_8$ and find the surprising result that the temperature range at which this gap becomes visible by angle-resolved photoemission spectroscopy is nearly an order of magnitude lower than the temperature range where the magnetic scattering becomes larger than the phonon scattering, as observed in the electrical resistivity measurements. Furthermore the spectral gap appears at temperature scales nearly an order of magnitude higher than the coherent temperature. We further show that when replacing In by Cd to tune the local density of states at the Ce$^{3+}$ site, there is a strong reduction of the hybridization strength, which in turn leads to the suppression of the hybridization gap at low temperatures.

\end{abstract}

\maketitle

Heavy fermion (HF) materials present properties of great interest related to non-trivial ground states, such as unconventional superconductivity and non-Fermi-liquid behavior, that frequently appear in the vicinity of a magnetically ordered state \cite{Review,Piers_JPCM_2001}. The microscopic nature of the hybridization between the 4$f$ and conduction electrons (which we label $ce$) is a key ingredient of the underlying physics of these phenomena. In some particularly cases,  the $f-ce$ hybridization at low-$T$ reveals a narrow stripe of spectral weight at zero frequency, which splits the conduction bands into two pieces separated by a hybridization gap \cite{Shim_Science}. As such, the existence of a temperature-dependent hybridization gap could be taken as a hallmark of HF behavior. 

The compound we study here, Ce$_2$RhIn$_8$ (Ce218), is a member of the Ce$_m M$In$_{3m+2}$ ($M$ = Co, Rh, Ir; and $m= 1, 2; $) family, which hosts many heavy fermions superconductors (HFS) in related structures \cite{Thompson_Fisk_review115}. Their structures have layers of \emph{M}In$_{2}$ stacked between \emph{m}-layers of CeIn$_3$ along the \textit{c}-axis, resulting in the 1-1-5 structure Ce$M$In$_5$ ($m = 1$) and the 2-1-8 structure Ce$_2M$In$_8$ ($m = 2$). Ce218 is an antiferromagnet  below 2.8 K, and under pressure, a HFS \cite{Nicklas}. Its crystal structure is shown in Fig. 1a. The electrical resistivity of Ce218  first decreases with decreasing temperature, but then increases below $\sim$ 200 K \cite{Thompson_Fisk_review115,Review} reaching a maximum at $T$ $\sim$ 5 K, below which the magnetic scattering becomes coherent. At lower $T$ a small kink is observed at $T_N$ = 2.8 K. The antiferromagnetic (AFM) transition can be tuned by exchanging In for Cd \cite{Pham_PRL2006,Cris_PRB2010}, as it can be seen in the magnetic specific heat data (C$_{mag}$) shown in Fig. 1b.  The figure shows that in Ce$_2$RhIn$_{7.79}$Cd$_{0.21}$ compounds, the substitution changes T$_N$ from 2.8 K to 4.8 K. 

In this work we report temperature dependent angle-resolved photoemission spectroscopy (ARPES) experiments on Ce218, which show the appearance of the hybridization gap in the electronic structure of the material at temperatures below $T$ = 20 K, and its suppression when the hybridization is tuned by doping Ce218 with Cd. We further find that the gap is maximum in the region close to the center of the 1st Brillouin zone (BZ), while being hardly detectable in the two outermost cylindrical sheets around the irreducible zone corner. The detailed evolution of the hybridization gap at low-$T$ is discussed in terms of the relevant energy scales of Ce218 in comparison to related HF materials.

The ARPES experiments were performed at the Synchrotron Radiation Center at the U1 4m-NIM beamline using photon energy of 22 eV, with a resolution of $\sim$ 15 meV and a Scienta R4000 spectrometer. Additional measurements at circular polarization were performed at the U9 VLS-PGM beamline with a Scienta 200U analyzer. The samples were cleaved \textit{in situ}, perpendicular to the {[}001{]} direction, and measured along cuts parallel to the [100] direction, mantaining a pressure below 5 x 10$^{-11}$ Torr.

Fig. 1c shows a sketch of the irreducible BZ. The Fermi surface sheets are obtained from a careful analysis of the momentum distribution curves at $E_F$ measured in Ce128. Detailed measurements of temperature, photon energy and polarization dependences are discussed in a related publication \cite{Rodolakis_2014}. Here we observe 3 electron-like sheets around the zone center and 4 barrel-like sheets centered at the corner. The dashed lines in Fig. 1c correspond to bands that are suppressed by matrix elements in the measurement configuration shown here. From extensive $k_z$ dispersion experiments as a function of photon energy~\cite{Rodolakis_2014}, we determined that at 22 eV photon energy, the ($k_x$ ,$k_y$)  =  (0,0) point in Fig. 1c is located at about 4/5 of the distance between $\Gamma$ and $Z$. Here we consider data along cuts in the BZ indicated by the dashed red lines in Fig 1c. In Fig. 1d we show the energy-momentum intensity map measured at 20~K and $h\nu$ = 22~eV along Cut 1 in Fig.1c. The small black-and-white inset (on the positive side of the momentum scale) was taken using circularly polarized incident photons to better show the band structure near $E_F$. We can clearly observe two bands in the vicinity of the Fermi energy: one displays a linear dispersion while the inner-most resembles an inverted parabola, which is just below the Fermi energy.

The details of the region enclosed by the white rectangle in Fig. 1d are shown in Fig. 2 as a function of temperature, for: a) $T = 200$ K; b) $T = 125$ K, c)  $T = 60$ K and d)  $T = 20$ K. 
Here the data are divided by a Fermi-Dirac (FD) distribution calculated using the experimental resolution and temperature. Although these intensity plots are illustrative, they do not provide quantitative information. More precise spectroscopic information  is obtained from the Energy Distribution Curves (EDCs) in Fig. 2e, which have also been divided by an effective FD distribution curve. The EDCs are obtained at momenta corresponding to the Fermi vector ($k_F$), \emph{i.e.} where the linear band (indicated by a dashed line in Fig. 2a) crosses the Fermi energy $E_F$. At $T =$ 200 K (Fig. 2a), the intensity map clearly shows that both bands described earlier cross $E_F$.
As the temperature is lowered to 125 K (Fig. 2b), the excitations develop flat-topped coherence peak above the Fermi energy (red dashed EDC in Fig. 2e). At 60K, more spectral weight is transferred below the Fermi energy, and once $T$ is lowered to 20 K, a small gap of about 4 meV develops. We assure ourselves of the presence of a gap by also comparing the raw EDC from the Ce218 sample at $T$ = 20 K  to that of a gold reference at the same chemical potential of the sample as shown in Fig. 2f. 

Strong hybridization effects are less prominent away from the zone center. Near the zone edge (Cut 2 in Fig. 1c) the barrel-like bands can be seen dispersing through $E_F$ in Fig. 2g at 60 K. At $T$ = 20 K, while the intensity map in Fig. 2h indicates a suppression of intensity at $E_F$ due to a sharpening of the  Fermi function, the FD divided EDCs (Fig. 2i) taken at positions  labelled $1$, $2$, and $3$ of Fig. 2h, indicate little hybridization effects. The EDCs show evidence for suppression of spectral weight for the inner-most barrels ($\#$1 and $\#$ 2 in Fig. 2i) but no spectral weight reduction is seen for the outer-most barrel ($\#$3 in Fig. 2i). Even though spectral weight reduction might occur, no clear spectral gap can be seen.

\begin{figure}
\centering
\includegraphics[width=1.0\columnwidth]{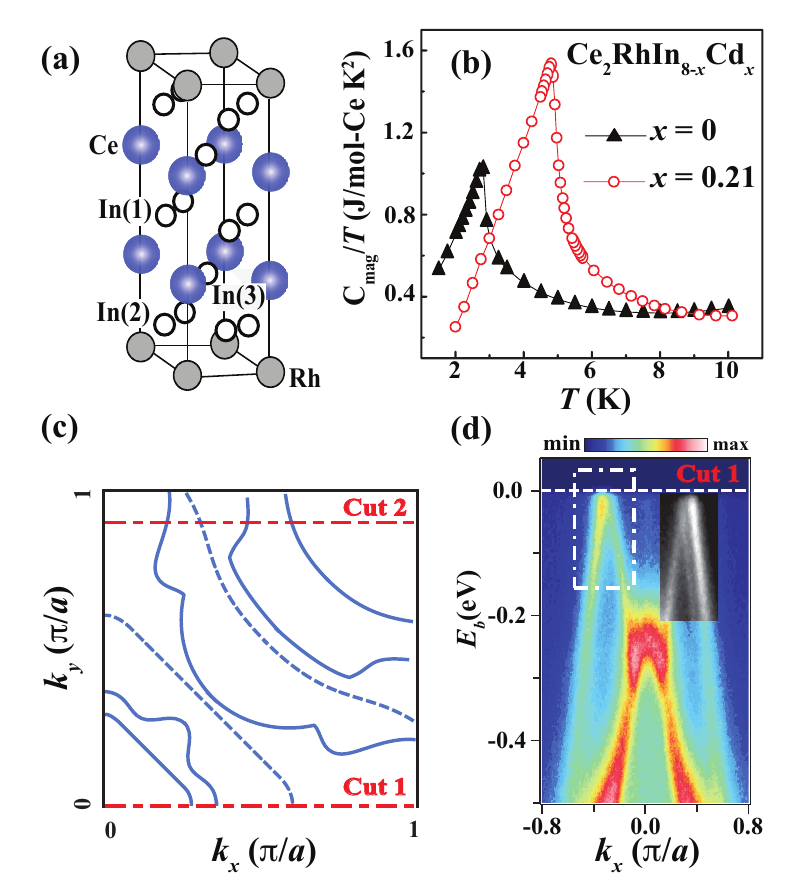}
\vspace{-0.9 cm}
\caption{a) Unit cell of Ce$_2$RhIn$_8$. b) C$_{mag}/T$ for pure ($x$ = 0) and Cd-doped ($x$ = 0.21) samples \cite{Cris_PRB2010}. The single crystals were grown from In-flux \cite{Nicklas}. c) Fermi Surface sheets of Ce$_2$RhIn$_8$ at low temperature~\cite{Rodolakis_2014}. The dashed lines correspond to bands that are suppressed by matrix elements in the configuration of the measurement presented here. The red dash-dotted lines represent the two regions of interest discussed in the text. d) Energy-momentum ARPES intensity map along a high symmetry direction, Cut 1 of \textbf{c)} taken at 20 K. The black-white region was taken  using circularly polarized incident photons.} \label{fig:Fig1}
\end{figure}

\begin{figure}
\centering
\includegraphics[width=1.0\columnwidth]{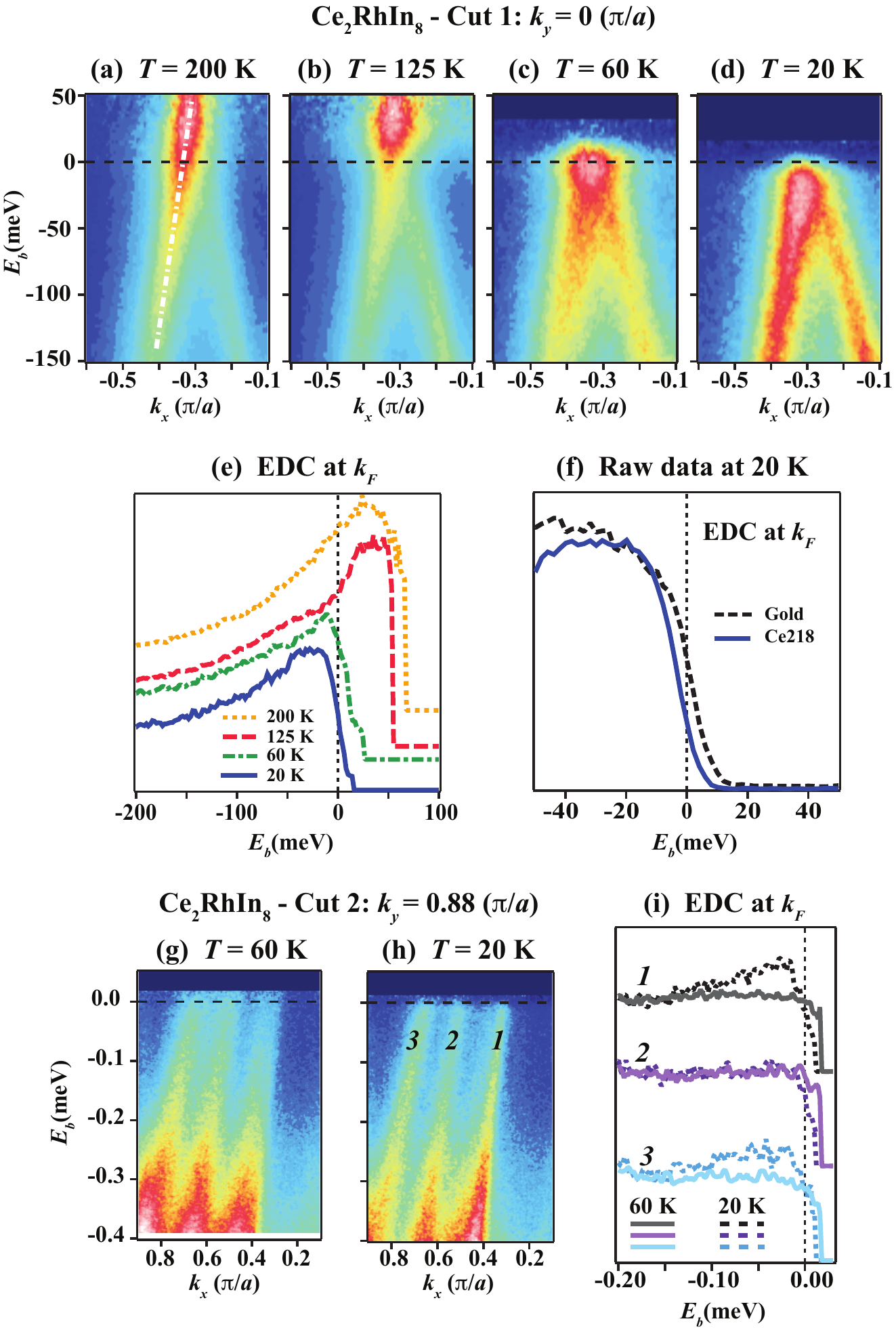}
\vspace{-0.7 cm}
\caption{ARPES intensity map of the conduction electrons near $E_F$ (white dash-dotted square of Fig.1d), measured at 22 eV and normalized by the FD distribution of each appropriate temperature for: a) 200 K; b) 125 K, c) 60 K nd d) 20 K. e) FD-divided EDCs taken at $k_F$ for the same temperature of the maps. f) Raw EDCs taken at 20 K comparing Ce218 to a gold reference measured at the same temperature. Energy-momentum ARPES intensity map of the conduction electrons along the Cut 2 of Fig. 1\textbf{c} measured at 22 eV at temperature: g) 60 K and h) 20 K. i) EDCs taken at $k_F$ for the positions labelled 1, 2 and 3 in panel h).}  \label{fig:Fig2}
\end{figure}

We also studied the effects of electronic tuning by replacing In by Cd (which contains one less $p$-electron). This substitution increases the magnetic specific heat, as shown in Fig. 1a, while suppressing Kondo scattering \cite{Cris_PRB2010}. Fig. 3 presents the ARPES data taken at 20 K for a Ce$_2$RhIn$_{7.79}$Cd$_{0.21}$ sample. Fig. 3a shows the FD-divided intensity map of the dispersion along Cut 1 of Fig 1c. Fig. 3b is an enlarged view of Fig. 3a, in the same energy interval used for Fig. 2c. Although the inverted parabolic band appears to be absent, in fact the Cd substitution results in a  shift in $E_F$. At this point, we cannot ascertain if there is a rigid band shift or not. The disappearance of the spectral gap can be more clearly seen in Fig. 3c, where we plot a comparison of the pure and doped compound EDCs at $k_F$  at $T = 20$ K. The doped compound, instead of a gap shows the appearance of a broad peak at low temperatures. 

Previously,  Raj \textit{et al}. \cite{Raj}  compared the ARPES measurements of Ce$_2$(Co,Rh)In$_8$ to 
LDA band calculations, arguing that the 4\textit{f} electrons are relatively more localized in Ce$_2$RhIn$_8$ than in Ce$_2$CoIn$_8$. Similar conclusions where reached by Fujimori \textit{et al}. \cite{Fujimori}, who found that the 4\textit{f} electrons are fully localized in CeRhIn$_5$ and CeIrIn$_5$. Koitzsch \textit{et al} \cite{Koitzsch} studied the temperature dependence of the dispersion and peak widths of CeCoIn$_5$ close to the zone corner of the BZ, suggesting the presence of some hybridization effects.  More recently Rui Jiang \textit{et al}. \cite{Rui_Jiang_ARPES_Ce218} compared ARPES data of the Ce$_2$RhIn$_8$ compound with density functional theory calculation, suggesting a scenario consistent with localization of the $f$ electrons. In contrast, our results clearly show the presence of a hybridization gap in a member of the Ce$_mM$In$_{3m+2}$ ($M$ = Co, Rh, Ir; and m = 1,2) family and its evolution with temperature and chemical substitution. 

We further note that the spectral gap is mainly observed in the states that exhibit $k_z$ dispersion, i.e. ones with 3D character. This suggests that the Ce 4\textit{f} states hybridize mostly with the out-of-plane $p$-states of In (from RhIn$_2$ layer). Our findings may be related to the theoretical and experimental results discussed by Shim \emph{et al}. \cite{Shim_Science} for the counterpart CeIrIn$_5$. The authors point out the presence of two values of the hybridization gap observed from optical conductivity depending whether the Ce is coupling with the in-plane or out-of-plane In $p$-electrons.

We now discuss the Cd-doping effects in Ce$_2$RhIn$_8$. Fig. 3a shows a lowering of the chemical potential, as the bands shift to lower binding energy (see Figs. 1d and 3a for comparison). We do not detect any obvious changes in the measured dispersions, so that the changes are akin to a rigid shift of the bands by  $\approx 0.1$ eV. Our ARPES data give microscopic evidence that the hybridization at the chemical potential is reduced upon lowering the $p$-electron count, shown by the suppression of the hybridization gap at $T$ = 20 K for the doped compound, visible in both the intensity map, Fig. 3b, and the EDC (light-green dotted curve in Fig. 3c). Now, a broad peak is clearly visible, together with a dip in the spectra. This shows the appearance of partial coherence, which in turn leads to the appearance of the spectral gap. We note that we have not considered a possible shift of one of crystal-field split levels of the $J = 5 / 2$ multiplet.

In fact, the suppression of the hybridization strength with Cd-substitution, along with a suggestion that the 4$f$-states hybridize mostly with the In out-of-plane $p$-states \cite{Shim_Science} as shown here, may explain why Cd-doping increases $T_{N}$ of the AFM phase for both tetragonal CeRhIn$_{5}$ and Ce$_{2}$RhIn$_{8}$ (\textit{m} = 2, \textit{n} = 1) \cite{Pham_PRL2006,Cris_PRB2010}, whereas it shows the opposite effect in the cubic CeIn$_{3}$ \cite{Berry_CdCeIn3_PRB2010}. In the CeIn$_{3}$
structure there is only one in-plane In(3) site which is weakly hybridized with the Ce$^{3+}$ 4$f$ electrons. As such, Cd-doping in this site does not significantly decrease the Kondo effect in favor of AFM. Therefore, other effects associated with Cd-doping, such as disorder and changes in the crystalline electric field (CEF) potential, can cause the suppression of $T_{N}$. Furthermore, the ground state tunability of the Ce$_{m}M$In$_{3m+2}$ compounds by changing $M$ = Co, Rh, Ir or Pd \cite{Thompson_Fisk_review115},  also fits with our findings: Ce $f$-electrons hybridize more strongly with out-of-plane In $p$-states which are sensitive to changes in $M$ and therefore these effects may play a role in the tuning between AFM and SC in this family.

To gain some qualitative insights from our results, we have modeled the $T$-crossover of the $f$-electron behavior as a function of the hybridization strength. We assume that the $f$-states are de-hybridized from the $ce$ bands at high-$T$, and evolve to a mixed state at low-$T$, giving rise to a coherent band of heavy quasi-particles. This \textit{de-hybridization} transition \cite{weger,weger1} can be approximately described using a Periodic Anderson Model where the many-body effects are
included through the finite lifetime $\tau_{f}$ of the correlated $f$-quasi-particles. The details of our calculation are shown in the Supplemental Information. The \textit{de-hybridization} transition is clearly exhibited through the total spectral density associated with electron excitations. The total spectral density $A(k_c, \omega)$ at the wave-vector $k_{c}$ where the original non-hybridized bands cross is shown in Fig. 4a, for temperatures above and below the \textit{de-hybridization} temperature $T_{V}$. For Ce218, we estimate $T_{V} $ to be $\approx 29$ K.

\begin{figure}[hbt]
\begin{center}
\vspace{-0.3 cm}
\includegraphics[width=0.9\columnwidth,keepaspectratio]{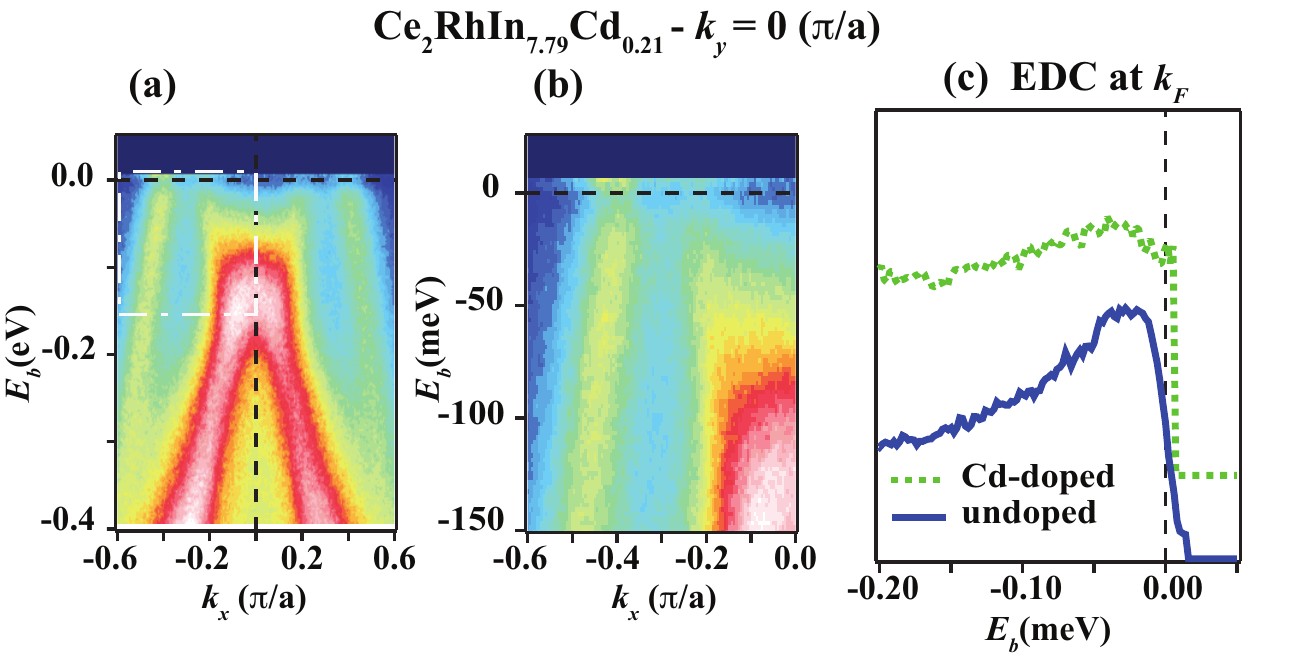}
\vspace{-0.5cm}
\end{center}
\caption{a) FD divided ARPES intensity map of the conduction electrons near $E_F$ measured at 20 K and 22 eV for Ce$_2$RhIn$_{7.79}$Cd$_{0.21}$. b) Enlarged view of the map of panel a) showing the equivalent energy range of Fig. 2c. c) FD normalized EDCs taken at $k_F$ to compare the pure and doped compounds at 20 K.}
\label{fig:Fig4}
\end{figure}

\begin{figure}
\begin{center}
\vspace{0.5 cm}
\includegraphics[width=0.7\columnwidth]{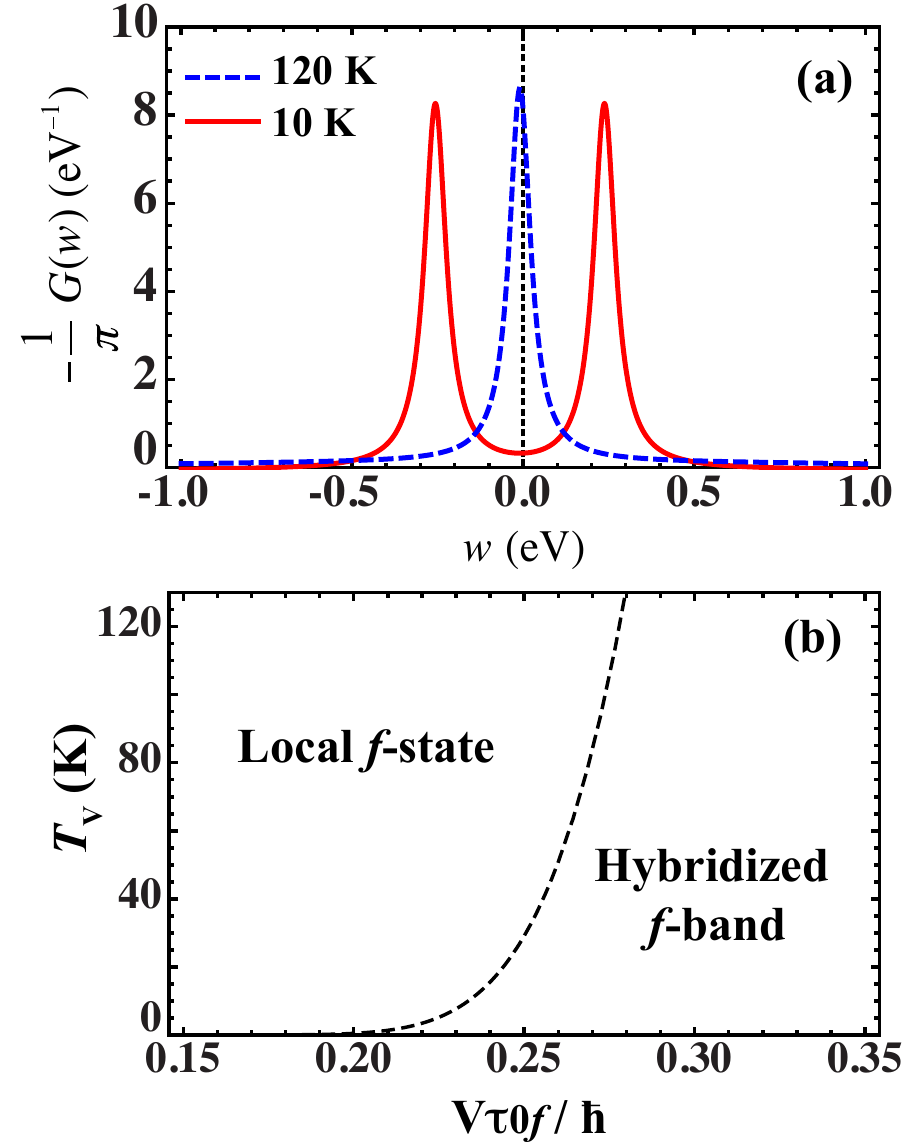}
\vspace{-0.3 cm}
\end{center}
\caption{a) The spectral density at $k = k_{c}$ for \textit{T} = 10 K (solid line) and \textit{T} = 120 K (dashed line). It was used, $E_{f}-\mu = -0.01$ eV, $V = 0.25$ eV, $\hbar\tau_{0f}^{-1} = 2$ eV and $T_{K} = 50$ K. b) $T_{V}$ as a function of $V\tau_{0f}/\hbar$. (See Supplemental Information).}
\label{Fig5}
\end{figure}

Fig. \ref{Fig5}b displays $T_{V}$  as a function of $V \tau_{0f}/\hbar$, where $V$ is the mixing term (i.e., hybridization) and $\tau_{f}$ is the finite lifetime of the correlated $f$-quasi-particles. It can be seen from Fig. \ref{Fig5}b that as $V$ or $\tau_{0f}$ decrease, $T_V$ drops. Above $T_V$, the $f$-local states remain decoupled from the $ce$ bands. In this \textit{de-hybridized} regime the $ce$ act essentially as carriers that mediate the RKKY (Ruderman-Kittel-Kasuya-Yosida) interaction between the local $f$-moments. In the regime of large $V$ and for $T<T_{V}$, the $f-ce$ electrons are strongly hybridized, giving rise to a narrow band of correlated quasi-particles. It then follows that a decrease in hybridization or of the lifetime of the $f$-states by, for example, doping the Ce218 with Cd, reduces $T_V$, consistent with our results. Further details on these possible effects are discussed in the Supplementary Material.

It is worthwhile noticing that Ce$_{2}$RhIn$_{8}$ seems to be propitious for the observation of the hybridization gap in our ARPES experiments, probably due to the energy scales of the problem. For Ce$_{2}$RhIn$_{8}$, the first CEF excited state is around T$_{CEF}$ $\sim$ 70 K \cite{arthur}, well above $T$ $\sim$ 5 K where the electronic coherent state actually fully develops.   Meanwhile for Ce$M$In$_{5}$ ($M$ = Co, Rh and Ir), $T_{CEF}$ $\sim$ 70 K \cite{CEF} is reasonably close to the maximum of the resistivity that appear about 50 K \cite{Thompson_Fisk_review115,Satoru,Nick}. Therefore, the fact that the hybridization effect will involve two CEF doublets with distinct hybridization strengths, and that presumably these compounds present higher coherence energy scale, may smother out the spectral features to a wider and higher temperature range  (See the Supplemental Information) which may make them difficult to be observed experimentally.

In conclusion, we report the observation of the hybridization gap in Ce$_2$RhIn$_8$ which is suppressed by Cd-doping. Furthermore our show strikingly distinct temperature scales related to the opening of the hybridization gap and to $f$-electron coherence. These separate energy scales for each phenomena suggest distinct hybridization strength involving the Ce$^{3+}$ 4$f$ CEF energy levels.

ARPES experiments were supported by UChicago Argonne, LLC, Operator of Argonne National Laboratory. Argonne, a US Department of Energy, Office of Science laboratory is operated under Contract DE-AC02-06CH11357. Z. F. thanks AFOSR MURI-USA; C.A., P.F.S.R. and P.G.P. thanks FAPESP (in particular grants No 2006/60440-0, 2009/09247-3, 2010/11949-3, 2011/01564-0, 2011/23650-5, 2012/04870-7), CNPq, and FINEP-Brazil; and M.A.C. acknowledges CNPq and FAPERJ for supporting this work.

%\bibliography{\string"basename of .bib file\string"}

\bibliography{basename of .bib file}

\end{document}